\documentclass[fleqn,12pt,twoside]{article}
\usepackage{espcrc1}

\usepackage{graphicx}
\usepackage[figuresright]{rotating}

\newcommand{\AmS}{{\protect\the\textfont2
  A\kern-.1667em\lower.5ex\hbox{M}\kern-.125emS}}

\title{Prospects in Classical Nova Modeling and Nucleosynthesis}

\author{Jordi Jos\'e\address[IEEC]{Departament de F\'\i sica i Enginyeria Nuclear, 
          Universitat Polit\`ecnica de Catalunya, and Institut d'Estudis 
	  Espacials de Catalunya (IEEC-UPC), Barcelona, Spain }}       

\begin{document}

\maketitle

\begin{abstract}

  Classical novae are fascinating stellar events, at the crossroads
  of astrophysics, nuclear physics and cosmochemistry. In this review,
  we outline the history of nova modeling with special emphasis on
  recent advances and perspectives in multidimensional simulations.
  Among the topics that are covered, we analyze the
  interplay between nova outbursts and the Galactic chemical abundances,
  the synthesis of radioactive nuclei of interest
  for gamma-ray astronomy, such as $^7$Li, $^{22}$Na or $^{26}$Al, 
  and the recent discovery of presolar meteoritic grains, 
  likely condensed in nova shells.

\end{abstract}

\section{Nuclear ashes: Classical novae and Galactic nucleosynthesis}

Classical novae are close binary systems consisting of a white dwarf, and a 
large main sequence (or a more evolved) star. The companion overfills its Roche lobe and 
matter flows through the inner Lagrangian point, leading to the formation of an
accretion disk around the compact star. A fraction of this (H-rich) matter
ultimately ends up on top of the white dwarf, where it is gradually 
compressed up to the point when ignition conditions to drive a
thermonuclear runaway (hereafter, TNR) are reached.

The thermonuclear origin of nova outbursts was first theorized  by
Schatzman \cite{Sch50,Sch51}. Modern multiwavelength
observations and numerical simulations (pioneered by the early hydro models of Starrfield et
al. \cite{Sta72}) have drawn a basic picture, usually referred to as the {\it thermonuclear runaway 
model}. 
Since then, several groups have attempted to improve our understanding of these dramatic
stellar events, including  state-of-the-art nova nucleosynthesis studies with spherically
symmetric (or 1-D) hydro codes (see \cite{KP97,Jos98,Sta98}, and references therein) or
preliminary multi-D approaches \cite{GL95,GLT97,KHT98,KHT99}.

Nuclear physics plays a crucial role in the course of the explosion.
As material from the accretion disk piles up on top of the (CO or ONe)
 white dwarf, the first nuclear reactions take place.
 This follows a rise in temperature since degenerate conditions unable
 the star to readjust the hydrostatic equilibrium by an envelope expansion
 and, as a result, a TNR ensues. The triggering reaction is 
 $^{12}$C(p,$\gamma$), which initiates the 'cold' CNO cycle.
 At very early stages of the explosion, the main nuclear activity
  is driven by  $^{12}$C(p, $\gamma$)$^{13}$N($\beta^+$)$^{13}$C(p, $\gamma$)$^{14}$N.
  But as the temperature rises, the characteristic time for proton capture reactions
  on $^{13}$N becomes shorther than its $\beta^+$ decay time, initiating the 'hot'
  CNO cycle. This is accompanyied by proton capture reactions onto $^{14}$N,
  leading to $^{15}$O, as well as by $^{16}$O(p,$\gamma$)$^{17}$F (near peak
  temperature, $T_{peak}$). At this stage, the envelope
  exhibits the presence of significant amounts of $^{13}$N, $^{14,15}$O and
  $^{17}$F. Indeed, it is the decay of these short-lived, $\beta^+$ unstable
  nuclei, carried away from the hot envelope's base to the outer,
  cooler layers by convective transport, what powers the ejection phase \cite{Sta72}:
  their sudden release of energy, few minutes after the peak of the explosion, increases
  the entropy and temperature of the material. As a result, the electron degeneracy
  is lifted and an overall expansion sets in, driving the 
  ejection of most (if not all) of the
  accreted envelope into the interstellar medium (indeed, the TNR is halted by the
   expansion of the envelope rather than by fuel consumption).
   The ejected layers enclose the history of multiple nuclear processes that
   modified its chemical composition during the course of the TNR: they are
  characterized by huge amounts of the daughther nuclei $^{15}$N, $^{17}$O, and
  to some extent, $^{13}$C, which have been claimed to represent the fingerprints of 
  classical nova outbursts
   in the overall Galactic history (\cite{KP97,Jos98,Sta98}, and references therein).
   Moreover, the envelope is also enriched in other species (depending
   on the nova type, the mass-accretion rate or the white dwarf initial luminosity),
   that may contribute to the Galactic abundances to a lesser extent. This includes
   nuclei such as $^{7}$Li, $^{19}$F, or $^{26}$Al.

   In order to match the energetics, peak luminosities, and associated nucleosynthesis of 
   the so-called {\it fast} novae, the modeling of the explosion requires
   mixing between the solar-like material transferred from the companion
   and the outermost layers (CO- or ONe-rich) of the underlying white dwarf. In fact, 
   at the typical temperatures expected during the course of the TNR, the amount
   of leakage from the CNO cycle is very limited \cite{Jos98}, and hence, the
   observed abundances of elements ranging from Ne to Ca (significantly overproduced with
   respect to solar proportions in some novae) cannot be explained in a natural way in terms
   of nuclear processes.  Indeed, 
   the  quest for a self-consistent mixing mechanism has become the {\it Holy Grail} of
   nova modeling:  several mechanisms have been proposed so far, including
   diffusion induced convection \cite{PK84,IFM91}, 
   shear mixing \cite{Mac83,LT87}, convective overshoot induced flame propagation \cite{Woo86}, 
   convection induced shear mixing \cite{KS89}, or more recently, 
    mixing by gravity wave breaking on white dwarf surfaces \cite{Ros01,Ale04}. But the
    final word has not yet been said...

 Among the species synthesized during classical nova outbursts, several
 radioactive nuclei have deserved special attention, in particular
 those associated with the predicted gamma-ray output from novae. Whereas
 $^{13}$N and $^{18}$F may be responsible for a prompt $\gamma$-ray
 emission at and below 511 keV,  
 $^7$Be ($^7$Li) and $^{22}$Na \cite{Gom98,Her99}, which decay much later, 
 may power line emission at 478 and 1275 keV, respectively. 
 $^{26}$Al is another interesting radioactive isotope that can be synthesized 
 during nova outbursts, 
 although due to its long lifetime only its cumulative emission can be observed.
 A detailed account of the predicted gamma-ray signatures of classical
 nova outbursts (including detectability distances) can be found in \cite{HJ04}.
 Hereafter, we will concentrate on the most relevant aspects of nova modeling,
  pointing out where do we stand                  
 and what is missing in our understanding of the overall phenomenon.
 In particular, we will focus on the synthesis paths
 of $^7$Li, $^{22}$Na and $^{26}$Al, the associated nucleosynthesis and observational
 constraints. 
 
 \subsection{$^7$Li}
 Recent hydrodynamic simulations 
 \cite{Her96,Jos98} have confirmed the feasibility of the {\it Be-transport mechanism} \cite{Cam55}
 as the key for $^7$Li production in nova outbursts:
 the process is initiated by the synthesis of $^7$Be through $^3$He($\alpha,\gamma$)$^7$Be,
 which is ultimately transformed into $^7$Li ($\tau \sim 77$ days) by means of an 
 electron capture, with the emission of a characteristic 478 keV $\gamma$-ray photon.
 Huge $^7$Li overproduction factors (i.e., $\sim 900$) with respect to solar values have been obtained,
 in particular for novae hosting CO white dwarf cores.
 These studies stressed the critical role played by the quasi-equilibrium between
 $^7$Be and $^8$B, driven by efficient photodisintegration reactions on $^8$B, on the survival
 of $^7$Be around peak temperatures, and confirmed novae as likely $^7$Li factories
 \cite{Sta78} (earlier hydrodynamic simulations assumed, however, envelopes in-place, thus neglecting 
 the possible impact of the initial stages of the TNR and the onset of convective 
 transport on $^7$Li production). Moreover, they refuted the conclusions based on simple
 1- and/or 2-zone models \cite{Bof93} that stressed the key role played by
 $^8$B(p,$\gamma$) in breaking the quasi-equilibrium between $^7$Be and $^8$B (and hence,
 leading to $^7$Be destruction instead).
 No nuclear uncertainties in the domain of nova temperatures significantly 
 affect $^7$Li synthesis \cite{Her96}.
 
 It is worth noting that the potential contribution of classical novae to 
 the Galactic $^7$Li content turns out to be small (i.e., less than 
 15\%, according to \cite{Her96,Jos98}). However, a nova contribution seems to be required to match 
 the $^7$Li content in realistic calculations of Galactic chemical evolution 
 \cite{Rom99}.

 \subsection{$^{22}$Na}
 The role of $^{22}$Na for diagnosis of nova outbursts was first outlined 
 in the seminal work of Clayton \& Hoyle \cite{Cla74}. This isotope decays into a short-lived 
 excited state of ${}^{22}$Ne, which de-excites to its ground state by emitting a $\gamma$-ray
 photon of $1.275$ MeV.  

 The synthesis of $^{22}$Na in novae proceeds through two alternative reaction
 paths. In the Ne-enriched envelopes of ONe novae \cite{Jos99}, it 
 takes place through $^{20}$Ne(p,$\gamma$)$^{21}$Na, followed either by 
 another proton capture and then, a $\beta^+$-decay into $^{22}$Na,  
 $^{21}$Na(p,$\gamma$)$^{22}$Mg($\beta^+$)$^{22}$Na, or decaying first into 
 $^{21}$Ne before another proton capture ensues, 
 $^{21}$Na($\beta^+$)$^{21}$Ne(p,$\gamma$)$^{22}$Na. The main destruction 
 channel at nova temperatures is $^{22}$Na(p,$\gamma$)$^{23}$Mg. 

 The nuclear uncertainties associated with the synthesis of $^{22}$Na in
 novae \cite{Jos99} have been recently reduced due to the first direct measurement of the 
 $^{21}$Na(p,$\gamma$) rate 
 with the DRAGON recoil separator at TRIUMF \cite{Bis03,Dau04}, 
 and to indirect determinations of the $^{22}$Na(p,$\gamma$) rate
 carried out with the Gammasphere at the Argonne National Lab 
 \cite{Jen04}.

\subsection{$^{26}$Al}
 $^{26}$Al was discovered in the interstellar medium by the HEAO-3
 satellite, through the detection of the 1809 keV $\gamma$-ray line.
  This characteristic $\gamma$-ray feature is
 produced by the $\beta^+$ decay 
 ($\tau = 1.04$ Myr) of the $^{26}$Al ground state to the first 
 excited state of $^{26}$Mg,
 which in turn de-excites to its ground state level by emitting a 1809 keV photon.

 The synthesis of $^{26}$Al requires moderate peak
 temperatures, of the order of $T_{peak} \leq 2 \times 10^8$ K,
 and a fast decline from maximum temperatures, conditions that are
 achieved in typical nova outbursts \cite{Pol95,Sta98,Jos99}.
 $^{26}$Al synthesis proceeds through 
 $^{24}$Mg(p, $\gamma$)$^{25}$Al($\beta^+$)$^{25}$Mg(p, $\gamma$)$^{26}$Al$^g$,
 whereas it is mainly destroyed by (p,$\gamma$) reactions
 \cite{Jos99}.
 A significant nuclear uncertainty affects the 
 $^{25}$Al(p,$\gamma$)$^{26}$Si rate \cite{Coc95,Jos99}, which translates into 
 an uncertainty in the expected contribution of novae to the Galactic 
 $^{26}$Al content. 
 Calculations based on recent prescriptions for the composition
 of ONe white dwarf cores
 suggest that the contribution of classical nova outbursts to the Galactic 
 $^{26}$Al abundance is small (i.e., less than 15\%), in agreement with the 
 results derived from the COMPTEL/CGRO map of the 1809 keV $^{26}$Al emission 
 in the Galaxy (see \cite{Die95}), which points towards young progenitors.

 \section{Observational constraints: from nova shells to presolar grains}

 The theoretical nucleosynthetic predictions described above can be compared with the
 abundance patterns inferred from observations of ejected nova shells (see 
  \cite{Jos98,Sta98}, and references therein).
 The comparison yields in
 general good agreement between models and observations:
 this includes atomic abundance determinations -H, He, C, O,
  Ne, Na...-, as well as a plausible endpoint for nova nucleosynthesis (around Ca), 
 suggesting that the thermal history of the explosion (i.e., $T_{peak}$, exposure times...) is
  reasonably well reproduced by current models. 
  Unfortunately, observations provide only direct information on atomic abundances and hence,
  do not pose severe constraints on the models. 

 Indeed, better constraints can be (partially) obtained 
 from the laboratory analysis of presolar grains, which
 yields isotopic abundance ratios. Classical novae are stellar dust factories:
  infrared measurements in a number of recent novae reveal the presence 
 of C-rich dust (Novae Aql 1995, V838 Her 1991, PW Vul 1984...),
 SiC (Novae Aql 1982, V842 Cen 1986...), hydrocarbons
 (Novae V842 Cen 1986, V705 Cas 1993...), or SiO$_2$ (Novae V1370 Aql 
 1982, V705 Cas 1993). Remarkable examples, such as novae QV Vul 1987,
 exhibit simultaneous formation of all those types of dust (see \cite{Geh98}
 for details on dust-forming novae). 
 Recently, several characteristic nova
 signatures have been identified by laboratory isotopic measurements in 
 five silicon carbide and two graphite grains isolated from the Murchison and Acfer 094 meteorites 
\cite{Ama01}. These tiny spherules, only a few microns in size, are characterized by low $^{12}$C/$^{13}$C
and $^{14}$N/$^{15}$N ratios,  $^{30}$Si excesses and close-to- or slightly 
lower-than-solar $^{29}$Si/$^{28}$Si ratios. In some cases,  high $^{26}$Al/$^{27}$Al 
 and low $^{20}$Ne/$^{22}$Ne ratios have been also determined. This discovery represents the first direct evidence
 of grains that exhibit nova signatures and opens up interesting possibilities for the future.

Theoretical efforts to predict the expected imprints of nova outbursts on
presolar grains have been conducted by different authors 
\cite{Sta97,Jos04}, including  preliminary estimates
on equilibrium condensation sequences in the ejected shells \cite{Jos04}:
these studies suggest that classical novae may contribute to the known
presolar corundum (Al$_2$O$_3$), spinel (MgAl$_2$O$_4$), enstatite
(MgSiO$_3$), silicon carbide (SiC) and silicon nitride (Si$_3$N$_4$)
grain populations.

\section{Multidimensional effects in nova outbursts}

The assumption of spherical symmetry in nova models (and in general, in stellar
explosions) excludes an entire sequence of events associated with the way in
which a TNR initiates (presumably as a point-source ignition) and propagates.
The first study of localized TNRs on white dwarfs was carried out by Shara \cite{Sha82} on
the basis of semianalytical models. He suggested that heat transport was too inefficient for a 
flame to spread a localized TNR to the rest of the white dwarf surface (i.e., the diffusively
propagated burning wave may require tens of years to extend throughout the whole stellar
surface).  Therefore,  he concluded that 
localized,  {\it volcanic-like} TNRs were likely to occur 
(mainly in $ M_{wd} \geq 1.2 M_\odot$ white dwarfs). 
But his analysis, based only on radiative and conductive transport, ignored the major 
(and crucial) role played by convection on the lateral thermalization of a TNR.

The importance of multidimensional effects for TNRs in thin stellar shells and surface layers 
(including classical nova outbursts) was revisited by Fryxell \& Woosley \cite{FW82}. 
They concluded that the most likely situation in nova outbursts involves TNRs propagated 
by small-scale turbulences. From dimensional analysis and flame theory, the authors derived a relation 
for the velocity of the deflagration front spreading
around the stellar surface: $v_{def} \sim (h_p v_{conv} / \tau_{burn})^{1/2}$,
where  $h_p$ is the pressure scale height and $v_{conv}$ the characteristic convective velocity.
Typical values for nova outbursts yield 
$v_{def} \sim 10^4$ cm s$^{-1}$ (that is, the flame propagates halfway through the stellar surface 
in about $\sim 1.3$ days).

The first, pioneering studies that addressed this issue in the framework of multidimensional hydro calculations
were performed by Shankar, Arnett \& Fryxell \cite{SAF92}, and Shankar \& Arnett \cite{SA94}.
 For that purpose, an accreting 1.25 $M_\odot$ white dwarf was evolved with
a 1-D hydro code and mapped into a 2-D domain (a spherical-polar grid of 25x60 km).
The explosive event was then followed with a 2-D version of the Eulerian code {\it PROMETHEUS}. 
A 12 isotope network, ranging from H to $^{17}$F, was included to account for the energetics of the
explosion.  Unfortunately, the subsonic nature of the
problem, coupled with the use of an explicit code (with a timestep limited by the Courant-Friedrichs-Levy
condition), posed severe limitations on the study, which was restricted to very 
extreme (rare) cases, characterized by huge T perturbations of about $\sim 100 - 600$\%, in small
local regions of the envelope's base. The overall computed time was also extremely small (about 1 second).
The calculations revealed that instantaneous, local temperature fluctuations cause Rayleigh-Taylor
instabilities. The rapid rise and subsequent expansion (in a dynamical timescale) cools down the
hot material and halts the lateral spread of the burning front, suggesting that such local temperature
fluctuations do not play a relevant role in the initiation of the TNR (in particular, at early stages).
The study, therefore, favors the occurrence of the local {\it volcanic-like} TNRs argued by \cite{Sha82}. 

Later on, Glasner \& Livne \cite{GL95}, and Glasner, Livne \& Truran \cite{GLT97},
extended these early attempts. New 2-D simulations  were performed with the code {\it VULCAN},
an arbitrarily Lagrangian Eulerian (ALE) hydrocode with capability to handle both explicit and implicit
steps. As in \cite{SAF92,SA94}, a slice of the star (0.1 $\pi^{rad}$), in spherical-polar coordinates 
with reflecting boundary conditions, was adopted.  The resolution near the envelope's base was around 5x5 km.
As before, the evolution of an accreting 1 $M_\odot$ CO white dwarf was initially followed by means of a 
1-D hydro code
(to overcome the early, computationally challenging phases of the TNR), and then mapped into a 2-D domain
as soon as the temperature at the envelope's  base reached $T_b \sim 10^8$ K.
As in \cite{SAF92,SA94}, the 2-D runs included a 12 isotope network.
The simulations revealed a good agreement with the {\it gross} picture described by 1-D models
(for instance, the critical role played by the $\beta^+$-unstable nuclei $^{13}$N, $^{14,15}$O, and
$^{17}$F, in the 
ejection stage, and consequently, the presence of large amounts of $^{13}$C, $^{15}$N and $^{17}$O in the ejecta).
 However, some remarkable differences were also found: first, the TNR was initiated by a myriad of
irregular, localized eruptions at the envelope's base caused by convection-driven temperature fluctuations. 
Hence, combustion proceeds as a chain of many localized flames (not as a thin front),
each surviving only a few seconds. Nevertheless, they concluded that turbulent diffusion 
efficiently dissipates any 
local burning around the core. As a result, they suggest that the fast stage of the TNR cannot be localized 
and therefore, the runaway {\it must} spread through the entire envelope.
Second, contrary to 1-D models, the core-envelope interface
is now convectively unstable, providing  a source for the envelope's metallicity enhancement 
 through a Kelvin-Helmholtz instability (a mechanism that bears a clear resemblance to 
the convective overshooting proposed by Woosley \cite{Woo86}). The efficient dredge-up of CO material from the
outermost white dwarf layers accounts for a $\sim 30$\% metal enrichment in the envelope
(the accreted envelope was assumed to be solar-like, without any arbitrary pre-enrichment prescription),
in agreement with the inferred metallicites in the nova shells ejected from CO novae.
And third, larger convective eddies, extending up to 2/3th of the envelope's height, with characteristic 
velocities of $v_{conv} \sim 10^7$ cm s$^{-1}$, were found in these 2-D simulations.
Nevertheless, and despite of these differences, the expansion and progress of
the TNR towards the outer envelope was almost spherically symmetric (although the initial burning
process was not). 

Results from other 2-D simulations were published, shortly after, by Kercek, Hillebrandt \& Truran \cite{KHT98},
with the aim to confirm the general features reported in \cite{GL95,GLT97}. In this case, a version
of the Eulerian PROMETHEUS code was used. A similar domain (a box of about 1800 x 1100 km) was adopted
despite a cartesian, plane-parallel geometry to allow the use of periodic boundary conditions, was chosen. 
Two resolution runs, one with a coarser 5x5 km grid (as in \cite{GLT97}) and a second with a finer 1x1 km grid, 
were performed.
Calculations used the same initial model than in Glasner et al. \cite{GLT97} and yield qualitatively
similar results but somewhat less violent outbursts (i.e., with longer TNRs and lower 
$T_{peak}$ and $v_{ejec}$), caused by large differences in the convective flow patterns:
whereas in \cite{GLT97}, a few, large convective eddies dominated the flow, most of the
 early TNR was now governed by small, very stable eddies
(with $l_{max} \sim$ 200 km) and, accordingly, more limited dredge-up and mixing episodes than in \cite{GLT97}
were found. The authors attributed such differences to the different
 geometry and, more significantly, the boundary conditions adopted in the simulations.

The only 3-D nova simulation to date has  also been performed by Kercek, Hillebrandt \& Truran \cite{KHT99}.
 The run, that adopted a computational domain of 1800x1800x1000 km, 
 with a resolution of  8x8x8 km, revealed flow patterns dramatically different 
 from those found in the 2-D simulations (much more erratic in the 3-D case):  mixing by turbulent 
 motions took place on very small scales (not fully resolved with the adopted resolution);
peak temperatures achieved were slightly lower than in
the 2-D case (a consequence of the slower and more limited dredge-up of core material). 
Moreover, the envelope attained a 
maximum velocity that was a factor $\sim 100$ smaller than the escape velocity and, presumably, 
no mass ejection 
was expected (except for a possible wind mass-loss phase). In view of these results, the authors
concluded that CO mixing {\it must} take place prior to the TNR, in contrast
with the main results reported by Glasner et al. \cite{GLT97}.

\section{To explote or not to explote: discussion and outlook}

 Despite the {\it thermonuclear runaway model} reproduces the gross observational
 features of classical nova outbursts, much remains to be done. The following list 
 (by no means exhaustive) outlines different aspects that require further attention:
 
- Identify the main mechanism (or mechanisms) responsible for mixing at the core/envelope
  interface.

- Reanalysis of the expected amount of material ejected per event, as compared with the values
  inferred from observations.

- Better spectra and analysis technics to provide theoreticians with reliable
   abundance patterns in nova shells.

- The observation of $^7$Li in the ejecta accompanying a nova outburst has been
 extraordinarily challenging. 
  Recently, the presence of this elusive isotope has been claimed for the first time:
 an observed feature compatible with the doublet at 6708 \AA \, of Li I has
 been reported in the spectra of V382 Vel (Nova Velorum 1999) \cite{Del02}.
  However, its has been
  argued that such observed feature in V382 Vel may correspond
  instead to another low-ionization emission centered at around 6705 \AA \,
  likely the doublet associated with N I \cite{Sho03}. Confirmation of the presence
  of this isotope in other nova shells would be highly desirable.

- $\gamma$-ray observations of specific gamma-ray signals: that would confirm another of
  the long-standing predictions of the {\it thermonuclear runaway model}, either by detecting line
  (478, 511, and 1275 keV) or continuum emission. 

- More theoretical and experimental work required to shed light into the contribution
    of novae to the Galactic $^{26}$Al content (including studies of nova frequencies and galactic 
    distributions)

- New nuclear physics experiments to reduce the uncertainties associated to key reactions
   (in particular, $^{25}$Al(p,$\gamma$) or $^{30}$P(p,$\gamma$)). 

- Identification of more nova grain candidates. In particular, it would be interesting
   to measure sulfur isotopic ratios (never achieved in presolar SiC grains) 
   and to identify as well nova candidates in the oxide grain population.

- More efforts in numerical modeling, specifically in the multidimensional framework.
    The limited (and somewhat contradictory) efforts to date have to be extended taking advantage
    of state-of-the-art, massive parallel architectures.
   First, it is crucial to understand the reason for the differences reported 
   by Glasner et al. \cite{GLT97} and Kercek et al. \cite{KHT98}: critical aspects, such
   as the specific geometry and the boundary conditions adopted, as well as
   numerical inaccuracies in the mapping procedure from 1-D models to a multi-D frame
   can certainly influence the outcome. Moreover, the different flow patterns found 
   when comparing 2- and 3-D simulations suggest that probably only 3-D modeling will bring
   final answers to our quest for the real nature of nova outbursts.

\vspace{0.5 cm}

I would like to thank M. Hernanz, S. Amari, A. Calder, A.E. Champagne, A. Coc, J. D'Auria, 
J. Dursi, C. Iliadis, S. Shore, 
S. Starrfield, J. Truran, M. Wiescher, and E. Zinner for many stimulating and enlightening discussions 
on several topics addressed in this manuscript.
 This work has been partially supported by the MCYT grants AYA2001-2360 and AYA2002-0494C03-01,  
and by the E.U. FEDER funds. Financial support from the catalan AGAUR during a sabbatical leave
is also acknowledged.

\end{document}